\documentclass[aps,prl,twocolumn,superscriptaddress,showpacs,amsmath,amssymb]{revtex4-1}

\usepackage{physics}
\usepackage{graphicx}
\usepackage[colorlinks=true,allcolors=blue]{hyperref}%

\begin{document}

\newcommand{\JLU}{Institut f\"ur Theoretische Physik,
  Justus-Liebig-Universit\"at, %Heinrich-Buff-Ring 16, 
  35392 Giessen, Germany}   
\newcommand{\HFHF}{Helmholtz Research Academy Hesse for FAIR (HFHF), Campus Giessen, 35392 Giessen, Germany}
\newcommand{\UB}{Fakult\"at f\"ur Physik, Universit\"at Bielefeld, D-33615 Bielefeld, Germany}

\title{Universal critical dynamics near the chiral phase transition and the QCD critical point} 

\author{Johannes V. Roth}
\affiliation{\JLU}

\author{Yunxin Ye}
\affiliation{\UB}

\author{S\"oren Schlichting}
\affiliation{\UB}

\author{Lorenz von Smekal}
\affiliation{\JLU}
\affiliation{\HFHF}

\date{September 2024}

\begin{abstract}
We use a novel real-time formulation of the functional renormalization group (FRG) for dynamical systems with reversible mode couplings to study Model~H, the conjectured dynamic universality class of the QCD critical point. We emphasize the structural similarities with Model~G, conjectured to be the dynamic universality class of the chiral phase transition in the limit of two massless quark flavors. Importantly, our formulation of the real-time FRG preserves all relevant symmetries throughout the FRG flow which, e.g., guarantees the non-renormalization of the reversible mode couplings, as but one exact result. We derive non-perturbative RG flow equations for the kinetic coefficients of both, Model~G and H, in parallel, and discuss commonalities and differences in the resulting fixed-point structure and dynamic critical exponents, such as weak-scaling relations which hold in either case versus the characteristic strong scaling of Model~G which is absent in Model H.
\end{abstract}

\maketitle

\paragraph{Introduction.}
Relativistic heavy-ion collision experiments at various beam energies allow to probe the phase structure of QCD at finite temperature and baryon density. One of the conjectured features of the QCD phase diagram thereby is the existence of a critical point at the end of a first-order transition line~\cite{Stephanov:1998dy}. To identify such a QCD critical point experimentally one searches for signatures of criticality, for example, in cumulants of conserved charges~\cite{Stephanov:2008qz}. However, since the fireball created in a heavy-ion collision is a rapidly evolving system, one needs to understand the time evolution of these critical fluctuations~\cite{Berdnikov:1999ph,Mukherjee:2015swa}. Similar to equilibrium/static critical phenomena, dynamic critical phenomena fall into universality classes~\cite{RevModPhys.49.435}. The basic idea is that one may be able to understand the dynamics of critical fluctuations in a heavy-ion collision by studying a simpler system from the same dynamic universality class.

The universality classes for dynamic critical phenomena were classified by Hohenberg and Halperin in Ref.~\cite{RevModPhys.49.435}. The ones relevant for critical dynamics in QCD, by their classification, are
Model~G of a four-component Heisenberg antiferromagnet, the dynamic universality class of the chiral transition in the two-flavour chiral limit \cite{Rajagopal:1992qz}, and Model~H of the liquid-gas transition in a pure fluid as well as the QCD critical point \cite{Son:2004iv}. 

To study critical dynamics naturally requires real-time methods. Classical-statistical simulations 
have been used to investigate the dynamic critical behavior of Models A, B, and C \cite{Berges:2009jz,Nahrgang:2018afz,Schweitzer:2020noq,Schweitzer:2021iqk,Schaefer:2022bfm,Chattopadhyay:2023jfm,Sieke:2024dns}, the $O(4)$ Model~G \cite{Schlichting:2019tbr,Florio:2021jlx,Florio:2023kmy}, as well as recently also Model~H \cite{Chattopadhyay:2024jlh}.
Functional methods such as real-time formulations of the functional renormalization group (FRG) can also be used to study dynamic critical behavior. Previous studies on critical dynamics using the real-time FRG mainly focused on the relaxational Models A, B, and C \cite{Canet:2006xu,Canet:2011wf,Mesterhazy:2013naa,Mesterhazy:2015uja,Roth:2023wbp}. Models~G and H, on the other hand, comprise the additional complication of non-linear reversible mode couplings which are protected from renormalization by Ward identities \cite{Roth:2024rbi}. Ignoring these subtleties, a first  real-time FRG study of Model~H led to reasonable values for the dynamic critical exponent of the order-parameter fluctuations \cite{Chen:2024lzz}. Because the truncation scheme was not symmetry preserving, however, additional ad hoc assumptions on the static FRG flow and on the reversible mode couplings had to be made to achieve these results.
In such a situation it is particularly important to have a formulation of the real-time FRG which manifestly preserves the non-trivial symmetries such as an extended temporal-gauge symmetry of the associated Martin-Siggia-Rose (MSR) action. In Ref.~\cite{Roth:2024rbi} we have demonstrated explicitly how this can be consistently achieved for Model~G, by introducing the external classical sources as well as the FRG regulators not directly in the MSR action but one level lower in the Landau-Ginzburg-Wilson (LGW) free energy of the system.
The same methodology applies to Model~H. 

In this paper we therefore review this real-time FRG formulation for Model~G and, in parallel, derive that for Model~H, emphasizing the structural similarities and differences between the two. In particular, our earlier work \cite{Roth:2024rbi} thus allows us to consistently formulate a symmetry-preserving real-time FRG flow  for the first time here for Model~H as well. As one of our main results, we derive non-perturbative RG flow equations for the kinetic coefficients of Models~G and~H. After introducing dimensionless combinations of the dynamic couplings, we obtain flow diagrams and study fixed-point structure, dynamic scaling relations, and the dependence of the resulting dynamic critical exponents on the number of spatial dimensions. Finally, we  comment on possible future applications. 

\paragraph{Equations of motion for Model G/H.} The static critical behaviour of an $N$-component order parameter $\phi$ is described by the Landau-Ginzburg-Wilson free energy,
\begin{equation}
    F =\!\int \! d^d x \left\{ \frac{1}{2}(\partial^i \phi_a)(\partial^i \phi_a)+\frac{m^2}{2} \phi_a \phi_a + \frac{\lambda}{4!N}(\phi_a\phi_a)^2 \right\} 
    \label{eq:freeEnergyPhi}
\end{equation}
in $d$ spatial dimensions, with $m^2<0$ to spontaneously break the $O(N)$ symmetry, and with quartic coupling~$\lambda$. To study \emph{dynamic} critical phenomena, one also needs to specify the equations of motion which reflect the
dynamic universality class of the system.
According to the classification by Halperin and Hohenberg \cite{RevModPhys.49.435}, this depends on whether the order parameter is conserved or not, whether there are other conserved quantities present in the system, and whether these exhibit non-trivial couplings with the order parameter, as expressed through non-vanishing Poisson-bracket relations.

In case of $O(N)$ Model~G \footnote{The generalization of the original ($N=3$)-component Model~G of Ref.~\cite{RevModPhys.49.435} to arbitrary $N$ is also known as the SSS model \cite{SASVARI1975108,taeuber_2014}.} the $N$-component order parameter is not conserved, but couples reversibly to a set of $N(N-1)/2$ conserved-charge densities, which can be summarized in an antisymmetric tensor $n_{ab}$. Statically, the fluctuations of the charge densities are non-critical and as such appropriately described by a quadratic term with susceptibility $\chi$ in the free energy, replacing
\begin{equation}
    F \to F + \int d^d x \, \frac{n_{ab} n_{ab}}{4\chi} \, .
\end{equation}
The charge densities and the order parameter satisfy the Poisson-bracket algebra
\begin{subequations}
\begin{align}
    \{\phi_a(\vec{x}), n_{bc}(\vec{x}')\} &=(\delta_{ac}\phi_b-\delta_{ab}\phi_c)\delta(\vec{x}-\vec{x}') \,,  \label{current2} \\
    \{n_{ab}(\vec{x}), n_{cd}(\vec{x}')\} &=(\delta_{ac}n_{bd}+\delta_{bd}n_{ac}\label{current3} \\  &\hspace{0.2cm} -\delta_{ad}n_{bc}  -\delta_{bc}n_{ad}) \delta(\vec{x}-\vec{x}') \, , \nonumber 
\end{align}
\end{subequations}
reflecting the fact that in Model~G the charge densities are the generators of $O(N)$ transformations. With these Poisson brackets, the equations of motion for Model~G can be written as~\cite{SASVARI1975108,taeuber_2014}
\begin{subequations}
\begin{align}
    \frac{\partial \phi_a}{\partial t} &= -\Gamma^{\phi} \frac{\delta F}{\delta \phi_a}+\frac{g}{2}\,\{\phi_a,n_{bc}\} \, \frac{\delta F}{\delta n_{bc}}+\theta_a \,, \label{eomop} \\
    \frac{\partial n_{ab}}{\partial t} &= \gamma \vec{\nabla}^2 \frac{\delta F}{\delta n_{ab}}+g\,\{n_{ab},\phi_c\} \, \frac{\delta F}{\delta \phi_c}+ \label{eomcc} \\ \nonumber &\hspace{4.0cm} \frac{g}{2}\,\{n_{ab},n_{cd}\} \, \frac{\delta F}{\delta n_{cd}}+\zeta_{ab}  \, ,
\end{align} \label{eq:eomsG}%
\end{subequations}
where the Gaussian white noises $\theta$ and $\vec{\zeta}$ have variances set by the fluctuation-dissipation relation (FDR), 
%\begin{subequations}
\begin{align}
    \langle \theta_a(t,\vec{x})\theta_b(t',\vec{x}') \rangle &= 2 \Gamma_{0} T \delta_{ab} \delta(\vec{x}-\vec{x}') \delta(t-t') \,, \label{eq:whiteNoise}\\
    \langle \zeta_{ab}(t,\vec{x}) \zeta_{cd}(t',\vec{x}') \rangle &= \nonumber\\
    &\hspace{-1.0cm} -2\gamma T (\delta_{ac}\delta_{bd}-\delta_{ad}\delta_{bc}) \vec{\nabla}^2  \delta(\vec{x}-\vec{x}') \delta(t-t')  \,,\nonumber
\end{align} %
%\end{subequations}
and drive the system towards the equilibrium Boltzmann distribution. For $N=4$, the $O(4)$ Model~G represents the dynamic universality class of the $SU(2)_L \times SU(2)_R \to SU(2)_V $ chiral phase transition in QCD with two massless quark flavors and anomalously broken axial $U_A(1) $ \cite{Rajagopal:1992qz}, where the six charge densities belong to the conserved iso-vector and iso-axial-vector currents.

In Model~H, the order parameter $\phi$ has one component, is conserved, and couples reversibly to the transverse part of the conserved momentum density $\vec{j}$ \cite{RevModPhys.49.435}. The static fluctuations of the latter are again non-critical. In the free energy \eqref{eq:freeEnergyPhi} the momentum density appears in the form of a kinetic energy with mass density $\rho$,
\begin{equation}
    F \to F + \int d^d x \; \frac{\vec{j}^{\,2}}{2\rho} \,.
\end{equation}
The components of the momentum density are the generators of spatial translations, so their Poisson-bracket algebra with the order parameter is given by~\cite{DZYALOSHINSKII198067}
\begin{align}
    \{\phi(\vec{x}),j_l(\vec{x}')\} &= \phi(\vec{x}')\frac{\partial}{\partial x'_l} \delta(\vec{x}-\vec{x}') \, ,\\
    \{j_l(\vec{x}),j_m(\vec{x}')\} &= \left[ j_l(\vec{x}') \frac{\partial}{\partial x'_m} \! - \! j_m(\vec{x}) \frac{\partial}{\partial x_l} \right] \! \delta(\vec{x}-\vec{x}') \, .\nonumber
\end{align}
With these Poisson brackets, we can express the equations of motion of Model~H \cite{PhysRevB.13.1299,RevModPhys.49.435} as
\begin{subequations}
\begin{align}
    \frac{\partial \phi}{\partial t}  &= \sigma \vec{\nabla}^2 \frac{\delta F}{\delta \phi} + g\{\phi,\vec{j}\} \cdot \frac{\delta F}{\delta \vec{j}} + \theta  \, ,\label{eomopH} \\
    \frac{\partial j_l}{\partial t} &= \mathcal{T}_{lm} \bigg[ \eta \vec{\nabla}^2 \frac{\delta F}{\delta j_m}+ g\{j_m,\phi\} \frac{\delta F}{\delta \phi} \;+ \label{eomojH} \\ \nonumber &\hspace{4.0cm} g\{j_m,j_n\}\frac{\delta F}{\delta j_n} \bigg]   + \xi_l \, ,
\end{align}\label{eq:eomsH}%
\end{subequations}
where $\mathcal{T}$ is a transverse projector, which in momentum space acts as $\mathcal{T}_{lm}(\vec{p}) = \delta_{lm} - p_l p_m/\vec{p}^2$. The Gaussian noise terms $\theta$ and $\vec{\xi}$ have variances
%\begin{subequations}
\begin{align}
    \langle \theta(t,\vec{x})\theta(t',\vec{x}')\rangle & =  -2 \sigma T \, \vec{\nabla}^2 \delta(\vec{x}-\vec{x}') \delta(t-t') \, ,\\
    \langle \xi_l(t,\vec{x})\xi_m(t',\vec{x}')\rangle & = -2 \eta T \, \mathcal{T}_{lm} \vec{\nabla}^2 \delta(\vec{x}-\vec{x}') \delta(t-t') \,.\nonumber
\end{align}
%\end{subequations}
Comparing the equations of motion of Model~G \eqref{eq:eomsG} and Model~H \eqref{eq:eomsH}, one immediately notices 
that the two merely differ by: (a) the number of components of the order parameter field, (b) its kinetic coefficient reflecting whether it is conserved (Model H) or not (Model G),  and (c) the detailed expressions for the reversible mode couplings. In Model~G, the term in \eqref{eomop} describes Larmor precession of the order parameter around the magnetic field generated by the fluctuating charge densities $n_{ab}$, while in Model~H, the corresponding term in \eqref{eomopH} describes advection of the order parameter with the momentum density $\vec{j}$. Other than that, the structural similarities ensure that the real-time FRG methods developed in \cite{Roth:2024rbi}   
can also be directly applied to Model~H.

\paragraph{Real-time FRG approach to Model G/H.} Non-per\-tur\-bative field-theoretic methods are needed to describe the critical behavior in the vicinity of a second-order thermal phase transition in $d=3$ spatial dimensions.
One such method which is particularly well approved for describing static critical phenomena is the Euclidean FRG, e.g., see~\cite{Litim:2001dt,DePolsi:2020pjk,DePolsi:2021cmi}. To describe \emph{dynamic} critical phenomena, one evidently needs real-time formulations, however. Since these phenomena are essentially classical, effective descriptions are usually based on the Martin-Siggia-Rose (MSR) formalism \cite{Martin:1973zz,Dominicis:1976,Janssen:1976}, where one introduces an auxiliary `response' field $\tilde{\psi}$ for every classical field $\psi$, which allows to express multi-time correlation functions as a path integral over an associated MSR action. 

By adding source terms of the form $J\tilde{\psi}$ and $\tilde{J}\psi$ directly to the MSR action, one can obtain a generating functional $Z[J,\tilde{J}]$ for multi-time correlation functions, following standard field theory techniques. In presence of reversible mode couplings, however, in order to obtain physical correlation functions \cite{PhysRevE.55.4120}, it was suggested in Ref.~\cite{Roth:2024rbi} to insert classical source terms of the form $J \psi$ in the free energy \eqref{eq:freeEnergyPhi} instead of adding $J \tilde{\psi}$ in the MSR action. Due to the non-trivial Poisson-brackets in the equations of motion \eqref{eq:eomsG} and \eqref{eq:eomsH}, this has the consequence that --  on the level of the MSR action -- these classical sources $J$ then couple to composite fields $\tilde{\Psi}$, which are related to the  standard response fields $\tilde{\psi}$ by a linear but field dependent transformation. We will therefore refer to those $\tilde{\Psi}$ as the `composite' response fields.

By coupling the sources at the level of the free-energy, i.e.~$F \to F - \int J \psi$, it is ensured that $(i)$ the classical symmetry of thermal equilibrium \cite{Sieberer:2015hba} as well as $(ii)$ an extended `temporal gauge' symmetry related to the reversible mode couplings, are maintained exactly in the presence of classical external sources. The extended symmetry $(ii)$ expresses an invariance under time-gauged $O(N)$ transformations in Model~G \cite{Roth:2024rbi}, and the invariance under time-gauged Galilean boosts in Model~H \cite{Canet:2022xnb}.

Next, to formulate the FRG flow equations \cite{Wetterich:1992yh}, 
one furthermore adds an RG-scale $k$ dependent (spatial) regulator term also on the level of the free energy, i.e.~$F \to F+\frac{1}{2}\int  \psi R_{k} \psi$, analogous to the way the (classical) external sources are introduced. The central object in the real-time generalization \cite{Berges:2012ty} of the standard FRG approach by Wetterich~\cite{Wetterich:1992yh} is the effective \emph{average} MSR action $\Gamma_k$, which is defined as the modified Legendre transform of the scale-dependent Schwinger functional $-i\log Z_k$, see Eq.~\eqref{eq:defGamk} in the Supplemental Material (SM). The RG-scale dependence of $\Gamma_k$ is described by an (exact) real-time flow equation \cite{Berges:2012ty}
\begin{equation}
    \partial_k \Gamma_k = -\frac{i}{2} \Tr\left\{ \frac{\partial_k R_k}{\Gamma_k^{(2)} - R_k } \right\} \, . \label{eq:treeLevelFlow}
\end{equation}
Based on the above symmetries $(i)$ and $(ii)$, it was shown in Ref.~\cite{Roth:2024rbi} that $(a)$ the mode-coupling constant $g$ is protected from renormalization, and $(b)$ the static FRG flow for the free energy \eqref{eq:freeEnergyPhi} is independent of the dynamics and simply given by the standard $d$-dimensional Euclidean flow equation \cite{Berges:2000ew}. In practice, these two exact results can largely simplify the derivation of flow equations. 

Introducing the source and regulator terms on the level of the MSR action in the standard way (as if there were no reversible mode couplings), and as done in   Ref.~\cite{Chen:2024lzz} also for Model H, the symmetries $(i)$ and $(ii)$ are generally not preserved. The violation of symmetries during the FRG flow, on the other hand, typically introduces RG relevant operators that would otherwise be excluded by symmetry.  These operators then need to be carefully fine-tuned to vanish in the IR again in order to reach the correct fixed point. To avoid such an additional fine tuning, and to guarantee the exact results that we discuss below, it is therefore necessary to set up an FRG flow that manifestly maintains all the relevant symmetries.

For illustration,  we first truncate the effective average MSR action by simply promoting all couplings ($m^2$, $\lambda$, $\Gamma^{\phi}$, and $\gamma$ for Model~G, and $m^2$, $\lambda$, $\sigma$, and $\eta$ for Model~H) to depend on the FRG scale~$k$. For the static FRG flow this amounts to a simple `$\phi^4$-truncation' which in combination with the flow of the kinetic coefficients here has the advantage to provide an entirely analytical description of the main qualitative features of dynamic scaling.
We employ the optimized regulator $R_k^{\phi}(\vec{p}) = (k^2-\vec{p}^{\,2})\theta(k^2-\vec{p}^{\,2})$ \cite{Litim:2001up} for the order parameter $\phi$ and set the regulators for the charge dentities $n_{ab}$ and $\vec{j}$ to zero, as their static fluctuations are anyway non-critical, and corresponding spatial regulator terms thus RG irrelevant.

The flow equations for the static quantities $m^2_k$ and $\lambda_k$ are given for arbitrary $N$ in Refs.~\cite{Berges:2000ew,Litim:2002cf,Roth:2024rbi}.
In addition, we use the FDR to project the flow onto the kinetic coefficients, as discussed for the order-parameter damping rate $\Gamma^{\phi}_k$ and charge mobility $\gamma_k$ of Model~G in Ref.~\cite{Roth:2024rbi}. For Model~H, the scale-dependent order-parameter mobility  $\sigma_k$ and shear viscosity $\eta_k$ can similarly by obtained from second functional derivatives of the effective average MSR action $\Gamma_k$ with respect to composite response fields $\tilde{\Phi}$ and $\tilde{\vec{J}}$ (cf.~Eq.~\eqref{eq:ModHResponseFields} in the SM) 
which are appropriate generalizations of the ones introduced for Model~G in Ref.~\cite{Roth:2024rbi},
\begin{align}
    \frac{1}{\sigma_k} &= \frac{1}{2iT} \lim_{\vec{p}\to 0} \vec{p}^{\,2} \lim_{\omega\to 0} \frac{\delta^2 \Gamma_k}{\delta \tilde{\Phi}(-p)\delta \tilde{\Phi}(p)} \bigg\rvert_{\text{0}} \, , \label{eq:defLambdak} \\
    \frac{1}{\eta_k} &= \frac{1}{2iT} \lim_{\vec{p}\to 0} \vec{p}^{\,2} \lim_{\omega\to 0} \frac{\mathcal{T}_{lm}(\vec{p})}{d-1} \frac{\delta^2 \Gamma_k}{\delta \tilde{J}_l(-p)\delta \tilde{J}_m(p)} \bigg\rvert_{\text{0}}\,,  \label{eq:defEtak}
\end{align}
with four momenta $p\equiv(\omega,\vec{p})$ and evaluated at vanishing field expectation values as indicated by $(\cdots)\rvert_{0}$.

Using the definitions in \eqref{eq:defLambdak} and \eqref{eq:defEtak}, or the corresponding ones for Model~G \cite{Roth:2024rbi}, one can project the flow equation \eqref{eq:treeLevelFlow} onto the various kinetic coefficients, see Eqs.~(\ref{eq:dsigmadk})--(\ref{Eq:dgammadk}) in the SM for the explicit expressions. The physically relevant information is then encoded in 
the following dimensionless combinations of the kinetic coefficients \cite{taeuber_2014}, first defined for Model~G,
\begin{align}
    w_G \equiv \chi \, \frac{\Gamma^{\phi}_k }{\gamma_k} \,, \hspace{0.5cm} f_G \equiv \frac{d\,\Omega_d\, g^2 T}{(2\pi)^d} \, \frac{k^{d-4}}{ \Gamma_k^{\phi} \gamma_k} \,, \label{definitionwfG}
\end{align}
and secondly for Model~H,
\begin{align}
    w_H \equiv \rho \, \frac{\sigma_k   k^2}{\eta_k} \,, \hspace{0.5cm} f_H \equiv \frac{d\,\Omega_d\, g^2 T}{(2\pi)^d} \, \frac{k^{d-4}}{\sigma_k \eta_k} \, .\label{definitionwfH}
\end{align}
Here, $w_G$ and $w_H$ are just the ratios between the kinetic coefficients of order parameter and conserved charges, while $f_G$ and $f_H$ conveniently represent the pre-factors of the loop diagrams in the flow of the kinetic coefficients.

\begin{figure}[t!]
    \centering
    \begin{minipage}{0.65\linewidth}
        \centering
        \textsf{\large \hspace{1.0cm} Model G}\\
        \includegraphics[width=0.95\linewidth]{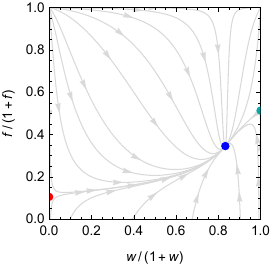}
    \end{minipage}
    \hfill
    \begin{minipage}{0.32\linewidth}
        \centering
        \phantom{\large Model G}
        \includegraphics[width=\linewidth]{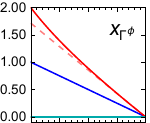}\\ \vspace{-0.04cm}
        \includegraphics[width=\linewidth]{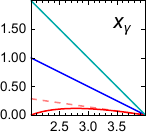}\\
    \textsf{\small spatial dimension \emph{d}}
    \end{minipage}  
    \caption{Left: flow diagram of $3d$ Model~G with one purely attractive stable strong-scaling fixed point (blue) at
    $(f^*_G,w^*_G) \approx (0.527,4.998)$ in our $\phi^4$-truncation,
    and two unstable weak-scaling fixed points at $w^*_G = \{ 0,\,\infty\}$ (red and green)
    \cite{taeuber_2014}. Right: the critical exponents  $x_{\Gamma^{\phi}}$ and $x_{\gamma}$ of order-parameter damping rate and charge mobility at the various fixed points (solid lines in matching colors), compared to the results from the  $1^\mathrm{st}$-order  $\epsilon$-expansion in Ref.~\cite{PhysRevE.55.4120} (dashed lines).}
    \label{fig:ModG}
\end{figure}

Inserting the usual dimensionless mass parameter 
$\bar{m}^2 \equiv k^{-2} m^2$
into the flow equations for the kinetic coefficients (cf.~Eqs.~(\ref{eq:dsigmadk})--(\ref{Eq:dgammadk}) in the SM), one obtains the dimensionless flow equations for $f_G$ and $w_G$ in Model~G,
\begin{align}
    k\partial_k f_G&= f_G (d-4) \;+  \label{eq:fGFlow} \\ \nonumber
    &\hspace{1.0cm} f_G^2 \left(\frac{2}{d(1+\bar{m}^2_k)^3}- (N-1) I_d(\bar{m}^2,w_G) \right) \, , \\
    k\partial_k w_G&= w_Gf_G \left[\frac{2}{d(1+\bar{m}^2_k)^3}\! + \! (N-1) I_d(\bar{m}^2,w_G) \right] \, , \nonumber
\end{align}
with the shorthand notation
\begin{align*}
    I_d(\bar{m}^2,w_G) &\equiv -\frac{1}{(1+\bar{m}^2)^2} \bigg\{ \frac{1}{1+(1+\bar{m}^2)w_G} \\ \nonumber 
    &\hspace{-1.2cm} + \frac{4-d}{d-2} \bigg[1 - {}_2F_{1}\left(1,\frac{d-2}{2}; \frac{d}{2}; -\frac{1}{(1+\bar{m}^2)w_G}\right) \bigg] \bigg\} \, .
\end{align*}
And the dimensionless flow equations for $f_H$ and $w_H$ in Model~H are obtained analogously as 
\begin{align}
    k\partial_k f_H &= (d-4) f_H\label{eq:fHFlow} + f_H^2\bigg[ \frac{1}{d(d+2)}\frac{1}{(1+\bar{m}^2)^3} \\ \nonumber
    &\hspace{-0.5cm} - \frac{2(d-1)}{d(d-2)}\bigg(\frac{1}{(1/w_H +(1+\bar{m}^2))^2}-\frac{1}{(1+\bar{m}^2)^2}\bigg)\bigg] \, ,\\ \nonumber
    k\partial_k w_H&= 2 w_H +w_H f_H\bigg[ \frac{1}{d(d+2)}\frac{1}{(1+\bar{m}^2)^3}  \label{eq:wHFlow}\\ \nonumber
    &\hspace{-0.5cm}+\frac{2}{d-2}\bigg(\frac{(d-1)}{d(1/w_H +(1+\bar{m}^2))^2}-\frac{d-1}{d(1+\bar{m}^2)^2} \bigg)\bigg]\,.
\end{align}
Importantly, Eqs.~\eqref{eq:fGFlow}--\eqref{eq:wHFlow} no-longer depend on the non-universal parameters $\rho$, $g$, and $T$, which stresses the universal character of these non-perturbative flow equations.

\begin{figure}
    \centering
    \begin{minipage}{0.65\linewidth}
        \centering
        \textsf{\large \hspace{1.0cm} Model H}\\
        \includegraphics[width=0.95\linewidth]{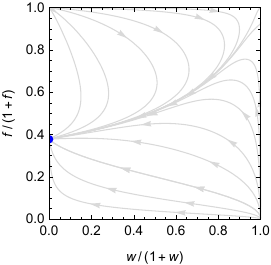}
    \end{minipage}
    \begin{minipage}{0.335\linewidth}
        \centering
        \phantom{\large Model H}
        \includegraphics[width=\linewidth]{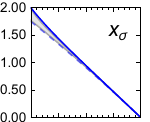}\\ \vspace{-0.13cm}
        \includegraphics[width=\linewidth]{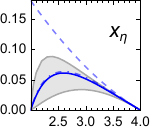}\\
    \textsf{\small spatial dimension \emph{d}}
    \end{minipage}
    \caption{Left: flow diagram of $3d$  Model~H with only a single stable weak-scaling fixed point at $w_H^*=0$. Right: the associated critical exponents $x_{\sigma}$ and $x_{\eta}$ of order-parameter mobility and shear viscosity in our $\phi^4$-truncation (solid blue lines) compared to the $2^\mathrm{nd}$-order $\epsilon$-expansion for Model~H in Refs.~\cite{Siggia:1976zz,Adzhemyan1999} (dashed blue), 
    and the result for $x_\eta$ (bottom) by Ohta and Kawasaki \cite{10.1143/PTP.55.1384} from mode-coupling theory (dashed-dotted blue). Gray bands between corresponding real-time FRG results from full LPA' truncations of the free energy in \eqref{eq:freeEnergyPhiN1LPA'}, based on the fixed expansion point $\rho_{0,k}=0$ (upper gray line) and on using the scale-dependent minimum $\rho_{0,k} = \rho_{\text{min},k}$ (lower gray line), are included to assess the quantitative uncertainty of the  $\phi^4$-truncation.}
    \label{fig:ModH}
\end{figure}

\paragraph{Dynamic critical behavior of Model G/H.} Close to a critical point, the dynamic critical exponent $z$ describes the divergence of the correlation time $\xi_t \sim \xi^{z}$ relative to the correlation length $\xi$. Sufficiently close to the critical point, on the other hand,  the relevant infrared cutoff during the FRG flow is set by the FRG scale $k>0$, and the correlation length effectively behaves as $\xi \sim 1/k$, while the correlation time $\xi_t$ can be inferred from the position $\omega_p$ of the lowest lying pole of the dynamic response function, as $\xi_t \sim 1/\text{Im}(\omega_p)$. Since for both, Model~G and H, the kinetic coefficients themselves show critical power-law divergences with the FRG scale~$k$, which can be quantified in terms of logarithmic $k$-derivatives of the kinetic coefficients in the scaling regime, i.e.~$x_{\Gamma^{\phi}} = -k\partial_k \log \Gamma^\phi_k$ and $x_{\gamma} = -k\partial_k \log \gamma_k$ for Model~G, or $x_{\sigma} = -k\partial_k \log \sigma_k$ and $x_{\eta} = -k\partial_k \log \eta_k$ for Model~H, these divergences also contribute to the dynamic critical exponents.

We first focus on the critical dynamics of the order parameter field  $\phi$. In Model~G the order parameter is not conserved, and an explicit calculation of its (retarded) propagator in our FRG truncation simply yields $\omega_{\phi} = -i \Gamma_k^\phi m_k^2 \sim k^{2-x_{\Gamma^{\phi}}}$, we therefore have $z_{\phi} = 2-x_{\Gamma^{\phi}}$. In contrast, in Model~H the order parameter, defined from the entropy per particle, is conserved, and one obtains the dispersion relation $\omega_{\phi} = -i\sigma_k \vec{p}^{\,2} (m_k^2+ \vec{p}^{\,2})$. Setting the external momentum to $|\vec{p}| \sim k$, one thus obtains an additional factor $k^2$ in the dispersion relation, yielding $\omega_{\phi}\sim \sigma_k  k^{4} \sim k^{4-x_{\sigma}}$, so that  $z_{\phi} = 4-x_{\sigma}$ in Model~H. 

Note that we have omitted the anomalous dimension $\eta_\perp$ of the order-parameter field here, which vanishes in our $\phi^4$-truncation of the spatial flow. For $\eta_\perp \neq 0$, these dynamic critical exponents generalize to $z_{\phi} = 2-\eta_\perp -x_{\Gamma^{\phi}}$ in Model~G, and $z_{\phi} = 4-\eta_\perp-x_{\sigma}$ in Model~H, respectively.

Moreover, in either case, the conserved charge densities $n_{ab}$ and $\vec{j}$ of Model~G and H, both also experience critical slowing down due to the reversible couplings to the order parameter. Setting the external momentum to $|\vec{p}| \sim k$ again, their dispersion relations, $\omega_{n} = -i \gamma_k \vec{p}^{\,2}/\chi$ and $\omega_{j} = -i\eta_k \vec{p}^{\,2}/\rho$, both  exhibit power laws,  $\omega_{n}\sim k^{2-x_{\gamma}}$ and $\omega_{j}\sim k^{2-x_{\eta}}$, as well, from which one can deduce the dynamic critical exponents $z_{n}= 2-x_{\gamma}$ for the charge densities $n_{ab}$ in Model~G and $z_{j}= 2-x_{\eta}$ for the transverse momentum density $\vec{j}$ in Model~H.

Since the critical point corresponds to a fixed point of the FRG flow, it is instructive to investigate the flow diagrams for Model~G and H, which are shown in Figs.~\ref{fig:ModG} and \ref{fig:ModH}, respectively, where the FRG flow lines are plotted in the compactified  $(w_{G/H},f_{G/H})$ plane \footnote{The dimensionless static coupling $\bar{m}^2$ is set to its value at the Wilson-Fisher fixed point within our truncation of the static free energy, which we obtain by solving the static sector of our flow equations. This poses no problem because the static flows are independent of the dynamics, as was explicitly shown in Ref.~\cite{Roth:2024rbi}. Moreover, we do not need the fixed-point value for the dimensionless quartic coupling $\bar{\lambda}$ here, since  the flow equations in (\ref{eq:fGFlow})--(\ref{eq:fHFlow}) for the dynamic couplings are independent of~$\lambda_k$ in our truncation, and hence only require scaling of $m_k^2$.}. One can see that for Model~G there are two unstable fixed points at $w_G^*=0$ and $w_G^*=\infty$. In addition, there is the stable fixed point that describes the critical dynamics of Model~G. In contrast, in the flow diagram of Model~H, there is only one stable fixed point at $w_H^*=0$. 
With this information we can then look for the stable fixed points in the dimensionless flow equations of $f_{G/H}$ analytically: 

For Model~G, with $w^*_G\not=0$ and $f_G^*\not=0$ in Eqs.~\eqref{eq:fGFlow}, the stable fixed point (in our truncation) is located at
\begin{align}
    f_G^*=\frac{(4-d)d(1+\bar{m}^2)^3}{4} \,.
\end{align}
For Model~H,  $w_H^*=0$ implies that 
the term with $1/w_H$ in the denominator in \eqref{eq:wHFlow} will tend to zero at the stable fixed point and thus can be safely neglected, yielding
\begin{align}
    f_H^*=\frac{(d-2) (4-d) (1+\bar{m}^2)^3}{\frac{2(d-1)}{d}(1+\bar{m}^2)+\frac{d-2}{d(d+2)}} \, . \label{ModelHf*}
\end{align}
Now, using  
$k\partial_k f_G=  
(d-4+\eta_\perp +x_{\Gamma^{\phi}}+x_\gamma) f_G$ 
for Model~G and $k\partial_k f_H= 
(d-4+\eta_\perp+x_\sigma+x_\eta) f_H$
for Model~H,  which follow directly from the definitions in Eqs.~\eqref{definitionwfG} and \eqref{definitionwfH} (here including field renormalization), we can deduce that as long as the fixed-point value $f^*$ is finite (i.e.~$0<f^*<\infty$), the following \emph{weak-scaling} relations \cite{taeuber_2014,PhysRevE.55.4120} hold at the fixed point in either case,
\begin{align}
    x_{\Gamma^{\phi}}+x_\gamma &= x_{\sigma}+x_\eta = 4-d  -\eta_\perp \, . \label{eq:weakScalingG}
\end{align}
If $w^*$ assumes a finite fixed-point value (i.e.~$0<w^*<\infty$) as well, as in the case of the stable fixed point in Model~G, then, from $k\partial_k w_G =  w_G (x_\gamma - x_{\Gamma^{\phi}}-\eta_\perp) $, we also have the \emph{strong-scaling} relation,
\begin{align}
    x_{\Gamma^{\phi}} = x_{\gamma} -\eta_\perp\,. \label{eq:strongScalingG}
\end{align}
With both $w_G^*$ and $f_G^*$ non-vanishing and finite at the stable strong-scaling fixed point of Model~G (cf.~Fig.~\ref{fig:ModG}), Eqs.~\eqref{eq:weakScalingG} and \eqref{eq:strongScalingG} are sufficient to uniquely fix the scaling exponents of both kinetic coefficients to
\begin{align}
    x_{\Gamma^{\phi}}+\eta_\perp  &=x_{\gamma}=2-\frac{d}{2} \,,
\end{align}
as represented by the solid blue lines for $x_{\Gamma^{\phi}}$ and $x_\gamma$  in the right panel of Fig.~\ref{fig:ModG}.
Moreover, with $ z_\phi=2-\eta_\perp -x_{\Gamma^{\phi}}$ and  $ z_n= 2 -x_\gamma $ from above, these dynamic scaling relations imply that one exactly recovers the dynamic critical exponents $z_\phi=z_n=d/2$ at this strong-scaling fixed point in Model~G. 

With $w_H^* = 0$ but $f_H^*\not=0 $ and finite at the Model~H fixed point (cf.~the left panel of Fig.~\ref{fig:ModH}), on the other hand, 
there is no such strong-scaling relation as in \eqref{eq:strongScalingG}. 
Instead, inserting Eq.~\eqref{ModelHf*} into the FRG flow equations for $\sigma_k$ and $\eta_k$ (cf.~Eqs.~\eqref{eq:dsigmadk} and \eqref{eq:detadk} in the SM), we then obtain the following analytical expressions for Model~H,
\begin{subequations}
\begin{align}
    x_{\sigma}&= \frac{2(d-1)f_H^*}{d\,(d-2)(1+\bar{m}^2)^2} \,,\\
    x_{\eta} &= \frac{f_H^*}{d\,(d+2)(1+\bar{m}^2)^3} \,,
\end{align}\label{eq:xSigmaAndxEta}%
\end{subequations}
with $f_H^*$ given by \eqref{ModelHf*},  for $\eta_\perp = 0 $ (the generalization for $\eta_\perp\not=0$ is given in the SM).
One can explicitly verify that an expansion of \eqref{eq:xSigmaAndxEta} in $\epsilon=4-d$ around $\epsilon=0$ reproduces the standard result from $1^\mathrm{st}$-order $\epsilon$-expansion \cite{Siggia:1976zz} exactly, but deviates at higher orders \footnote{We emphasize that our results are not restricted to the presumably small regime around $d=4$ in which $1^\mathrm{st}$-order $\epsilon$-expansion is valid. On the contrary, the power of non-perturbative/self-consistent approaches is that they are capable of describing the correct qualitative behaviour of certain observables for large values of the expansion parameter ($\epsilon \sim 1-2$ here), even if they deviate from exact perturbative results for small values of the expansion parameter ($\epsilon \ll 1$ here).}. Our FRG results \eqref{eq:xSigmaAndxEta} are plotted as the solid blue lines in the right panel of Fig.~\ref{fig:ModH} along with results from $2^\mathrm{nd}$-order $\epsilon$-expansion \cite{Siggia:1976zz,Adzhemyan1999}, which currently is the highest order available in this case.
While the critical exponent of the  mobility $x_\sigma $ deviates from the latter only on a quantitative level, that of the shear viscosity shows an entirely different qualitative behavior. In fact, our FRG result for $x_\eta$ in the $\phi^4$-truncation  closely resembles that of a self-consistent mode-coupling calculation with $\eta_\perp =0 $ by Ohta and Kawasaki \cite{10.1143/PTP.55.1384}.
Specifically, with $z_\phi = d+x_\eta $,  in $d=3$ spatial dimensions, we obtain the dynamic critical exponent $z_\phi \approx 3.051$ while the result from \cite{10.1143/PTP.55.1384} corresponds to $z_\phi \approx 3.054 $; for comparison, the $2^\mathrm{nd}$-order $\epsilon$-expansion yields
$z_\phi \approx 3.071$ \cite{Adzhemyan1999}.

In order to verify the robustness of our analytical results for Model H from this simple $ \phi^4$-truncation, we also consider 
the extended local-potential approximation (LPA') for the static free energy \eqref{eq:freeEnergyPhi}, using the Ansatz
\begin{align}
    F_k =\!\int \! d^d x \left\{ \frac{Z_k^{\perp}}{2}(\vec{\nabla}\phi)^2 +U_k(\rho) \right\}\;, \quad \rho=\phi^2\;. 
    \label{eq:freeEnergyPhiN1LPA'}
\end{align}
This truncation includes the full field dependence of the effective potential $U_k(\rho)$, and, in addition, a non-trivial wave function renormalization factor $Z_k^{\perp}$ evaluated at some (possibly $k$-dependent) field expansion point $\rho_{0,k}$, which gives rise to a non-vanishing anomalous dimension $\eta_{\perp} = -k\partial_k \log Z_k^\perp$. 

For the details of this LPA' truncation we refer to the SM, where the static flow equations for $U_k(\rho)$ and $Z_k^{\perp}$ are given in Eqs.~\eqref{eq:dkUkLPA'} and \eqref{eq:anomDimLPA'}, and the flow equations for the kinetic coefficients of Model~H with only slight modifications for $\eta_\perp\not=0$ in Eqs.~\eqref{eq:dsigmadkEtap} and \eqref{eq:detadkEtap}. We consider two versions of the LPA', where the flow equations for $Z_{k}^{\perp}$, $\sigma_{k}$ and $\eta_{k}$ are evaluated either at vanishing field expectation value,  $\rho_{0,k}\equiv 0$ \footnote{Note that with $\rho_{0,k} \equiv 0$ one recovers standard LPA again, since the flow $k\partial_k Z_k^{\perp} = 0$ vanishes in this case.}, or at the scale dependent minimum of the effective potential, $\rho_{0,k}=\rho_{\text{min},k}$.
The numerical results from the LPA' are shown as solid gray lines in the right panel of Fig.~\ref{fig:ModH}. For both diagrams ($x_{\sigma}$ and $x_{\eta}$) the upper gray curves correspond to $\rho_{0,k} \equiv 0$, and the lower gray curves to using the comoving $\rho_{0,k} =\rho_{\text{min},k}$, which includes a non-vanishing anomalous dimension $\eta_{\perp} \neq 0$. One can see that the qualitative behavior of the $\phi^4$-truncation remains unchanged, in particular with $x_{\eta}\to 0$ for $d \to 2$. The corresponding numerical values for the dynamic critical exponent in $d=3$ are $z_{\phi} \approx 3.058$ for fixed  $\rho_{0,k}=0$,
and $z_{\phi} \approx 3.034$ for the comoving $\rho_{0,k}=\rho_{{\rm min},k}$ with $\eta_\perp \not=0$ \footnote{For comparison, the result from the corresponding comoving LPA (with $\eta_\perp =0$) is practically indistinguishable, yielding $z_\phi \approx 3.033$ in $d=3$, which shows that the effect of field renormalization is negligible here.}.   Since an exact solution of the FRG flow would be independent of the expansion point $\rho_{0,k}$, these residual differences can serve to estimate the systematic uncertainties in the truncation.

\paragraph{Summary \& Outlook.}
We have used our novel real-time FRG approach to dynamical systems with reversible mode couplings from Ref.~\cite{Roth:2024rbi} to study, in parallel, the critical dynamics of Model~G and Model~H, the conjectured dynamic universality classes of the chiral transition in QCD with two flavors of nearly  massless quark and of the QCD critical point at finite baryon density, respectively. Carefully adding external sources and regulators at the level of the LGW free energy, this formalism ensures the preservation of all relevant symmetries and, in particular, provides Ward identities to protect the reversible mode coupling constant $g$ from renormalization. 

Within this framework we have derived one-loop exact non-perturbative FRG flow equations for the kinetic coefficients of Model~G and H. 
The associated fixed-point structure is qualitatively in line with $1^{\mathrm{st}}$-order $\epsilon$-expansion \cite{taeuber_2014}, comprising one stable strong-scaling and two unstable weak-scaling fixed points in Model~G, as compared to only a single weak-scaling fixed point in Model~H. We have obtained the corresponding strong and weak dynamic relations and computed the dynamic critical exponents $x_i$ of the kinetic coefficients in $2<d<4$ spatial dimensions. 
For the $O(4)$ chiral transition in QCD these determine the singular behavior of chiral order-parameter and Goldstone pion damping rate $\Gamma^\phi_k \sim k^{-x_{\Gamma^\phi}}$, and the iso-(axial-)vector charge mobilities $\gamma_k \sim k^{-x_\gamma}$ of Model~G. The mobility $\sigma_k \sim k^{-x_\sigma} $ of the conserved order parameter of Model~H determines the singular behavior of the entropy per baryon at the QCD critical point and that of transverse momentum densities the shear viscosity $\eta_k \sim k^{-x_\eta} $. For the dynamic critical exponents $z_i$ the strong-scaling relations of Model~G imply the exact result $z_\phi = z_n = d/2$. In Model~H, we obtain numerically in $d=3$ spatial dimensions $x_\sigma \approx 0.949$ and $x_\eta \approx 0.051 $ in quite close agreement with the mode-coupling calculations by Ohta and Kawasaki \cite{10.1143/PTP.55.1384}, implying  $z_\phi \approx 3.051 $ for the critical fluctuations of the entropy per baryon.

Evidently, the true power of this formalism is that it can straightforwardly be extended to include more sophisticated truncations of the effective action, e.g.~improving the static truncation for the effective LGW free energy, including derivative couplings and anomalous dimensions, for improved precision. Moreover,
including self-consistent frequency and/or momentum dependencies of the kinetic coefficients $\sigma_k(\omega,\vec{p})$ and $\eta_k(\omega,\vec{p})$, cf.~Refs.~\cite{Roth:2023wbp,Roth:2024rbi}, the framework can be used to compute universal dynamic scaling functions \cite{Schweitzer:2020noq,Schweitzer:2021iqk,Roth:2024rbi} that go beyond the common Kawasaki approximation \cite{Kawasaki:1970dpc,Stephanov:2017ghc,Pradeep:2022mkf}.

The developments presented in this letter also provide the basis for  applications of this real-time FRG approach to various other systems with reversible mode couplings. One example is stochastic hydrodynamics,  where the non-linear self-coupling of the momentum density $\vec{j}$ (cf.~Eq.~\eqref{eomojH}) leads to a non-trival renormalization of the shear viscosity~\cite{Kovtun:2011np}. By using the present techniques, this effect could be computed non-perturbatively and compared to the one-loop result of Ref.~\cite{Kovtun:2011np}.

\paragraph{Acknowledgements.}
We thank Adrien Florio, Eduardo Grossi, Fabian Rennecke, Alexander Soloviev, and Derek Teaney for insightful discussions. This work was supported by the Deutsche Forschungsgemeinschaft (DFG, German Research Foundation) through the CRC-TR 211 ‘Strong-interaction matter under extreme conditions’-project number 315477589 – TRR 211. JVR is supported by the Studienstiftung des deutschen Volkes. 

\bibliographystyle{h-physrev3}
\bibliography{refs}

\newpage

\renewcommand{\thesubsection}{{S.\arabic{subsection}}}
\setcounter{section}{0}

\onecolumngrid

\section*{Supplemental Material}

In this supplement we illustrate in more detail how one translates the real-time FRG formalism for Model~G developed in Ref.~\cite{Roth:2024rbi} to Model~H. The stochastic equations of motion \eqref{eq:eomsH} of Model~H are described by the MSR partition function
\vspace{-.4cm}
\begin{align}
    Z = \int \mathcal{D}\phi \mathcal{D}\tilde{\phi} \mathcal{D}j \mathcal{D}\tilde{j}\,e^{iS} = 1 \, , \label{eq:partFnc}
\end{align}
with MSR action
\vspace{-.4cm}
\begin{align}
    S = \int_x \bigg\{ &-\tilde{\phi}\left(\frac{\partial \phi}{\partial t} - \sigma \vec{\nabla}^2 \frac{\delta F}{\delta \phi} - g\{\phi,\vec{j}\} \cdot \frac{\delta F}{\delta \vec{j}} \right) \nonumber \\
    &-\tilde{j}_l \left( \frac{\partial j_l}{\partial t} - \mathcal{T}_{lm} \bigg[ \eta \vec{\nabla}^2 \frac{\delta F}{\delta j_m}+ g\{j_m,\phi\} \frac{\delta F}{\delta \phi} +  g\{j_m,j_n\}\frac{\delta F}{\delta j_n} \bigg] \right) \nonumber \\
    &- iT \tilde{\phi} \sigma\vec{\nabla}^2 \tilde{\phi} - iT  \tilde{j}_l \eta \mathcal{T}_{lm} \vec{\nabla}^2 \tilde{j}_m \bigg\} \, . \label{eq:MSRAction}
\end{align}
As explained in the main text, by adding physical source terms to the free energy $F \to F - J\psi$ (instead of adding $S \to S + J\tilde{\psi}$ to the MSR action), and unphysical source terms to the MSR action $S \to S + \tilde{J}\psi$ as usual, the partition function \eqref{eq:partFnc} is promoted to a generating functional for multi-time correlation functions,
\begin{align}
    Z[H,\tilde{H},\vec{A},\tilde{\vec{A}}] &= \int \mathcal{D}\phi\,\mathcal{D}\tilde{\phi}\,\mathcal{D}n\,\mathcal{D}\tilde{n}\, \exp\bigg\{ iS + i\int_x \big( \tilde{H} \phi + \tilde{A}_l j_l \big) \; +  \label{eq:genFunc} \\ \nonumber &\hspace{-1.0cm} i\int_x H\big( -\sigma \vec{\nabla}^2 \tilde{\phi} + g\{\phi,j_m\} \mathcal{T}_{mo} \tilde{j}_o \big) + i\int_x  A_l \big(-\eta \vec{\nabla}^2 \mathcal{T}_{lo} \tilde{j}_o + g \mathcal{T}_{lm}\{j_m,\phi\} \tilde{\phi}  + g \mathcal{T}_{lm} \{j_m,j_n\} \mathcal{T}_{no} \tilde{j}_o \big) \bigg\} \, .
\end{align}
We see from Eq.~\eqref{eq:genFunc} that the physical sources $H$ and $\vec{A}$ couple to the following composite operators:
\begin{subequations}
\begin{align}
    \tilde{\Phi} &\equiv -\sigma \vec{\nabla}^2 \tilde{\phi} + g\{\phi,j_m\} \mathcal{T}_{mo} \tilde{j}_o \, , \\
    \tilde{J}_l &\equiv   -\eta \vec{\nabla}^2 \mathcal{T}_{lo} \tilde{j}_o + g \mathcal{T}_{lm}\{j_m,\phi\} \tilde{\phi} + g \mathcal{T}_{lm} \{j_m,j_n\} \mathcal{T}_{no} \tilde{j}_o \, .
\end{align} \label{eq:ModHResponseFields}%
\end{subequations}
Since these are proportional to the standard MSR response fields $\tilde{\phi}$ and $\tilde{\vec{j}}$, we refer to $\tilde{\Phi}$ and $\tilde{\vec{J}}$ as `composite' response fields (the transverse projectors $\mathcal{T}$ in \eqref{eq:ModHResponseFields} ensure that only the transverse parts of the momentum densities contribute).

Adding regulator terms of the form $F \to F + \frac{1}{2}\psi R_k \psi$ to the free energy, the generating functional \eqref{eq:genFunc} becomes dependent on the FRG scale~$k$, $Z \to Z_k$. 
As in the standard FRG approach by Wetterich \cite{Wetterich:1992yh}, we define the effective average MSR action $\Gamma_k$ as the modified Legendre transform of the associated scale-dependent Schwinger functional $-i\log Z_k$,
\vspace{-.4cm}
\begin{align}
    \Gamma_k[\phi,\tilde{\Phi},\vec{j},\tilde{\vec{J}}] \equiv \sup_{H,\tilde{H},\vec{A},\tilde{\vec{A}}}\big\{ -i \log Z_k[H,\tilde{H},\vec{A},\tilde{A}] - \int_x \big( \tilde{H}\phi + H \tilde{\Phi} + \tilde{A}_l j_l + A_l \tilde{J}_l \big) \big\} \, , \label{eq:defGamk}
\end{align}
and $\Gamma_k$ then satisfies the (exact) real-time flow equation \eqref{eq:treeLevelFlow} \cite{Berges:2012ty}. 

For the scope of this work, our truncation of the full flow equation \eqref{eq:treeLevelFlow} is to simply promote the couplings appearing already in the bare MSR action \eqref{eq:MSRAction} to depend on the FRG scale~$k$, i.e.~$m^2 \to m_k^2$, $\lambda \to \lambda_k$, $\sigma \to \sigma_k$, and $\eta \to \eta_k$. This allows for an analytical treatment of the flow equations. 

The full propagator $G_k$ is related to the inverse of the 2-point function, $G_k = -(\Gamma_k^{(2)}-R_k)^{-1}$, where $\Gamma_k^{(2)}$ denotes the second functional derivative of $\Gamma_k$. In particular, the dynamic retarded/advanced response functions (at vanishing field expectation values) read (in the notation of Ref.~\cite{Roth:2024rbi})
\vspace{-.2cm}
\begin{align}
    G^{R/A}_{\phi,k}(\omega,\vec{p})
    &= -\frac{\sigma_k \vec{p}^2}{\pm i\omega-\sigma_k \vec{p}^2(m_k^2 + \vec{p}^2 + R_k^{\phi}(\vec{p}) )} \,, \hspace{1.0cm}
    G^{R/A}_{j,k}(\omega,\vec{p}) = -\frac{\eta_k \vec{p}^2}{\pm i\omega-\eta_k \vec{p}^2(1/\rho + R_k^{j}(\vec{p}) )} \,. \label{eq:ModHGRandGA}
\end{align}
Since the symmetry of thermal equilibrium is preserved during the FRG flow, the statistical functions are set by the classical FDR,
\vspace{-.4cm}
\begin{align}
    iF_{\phi/j,k}(\omega,\vec{p}) &= \frac{T}{\omega} \left( G^R_{\phi/j,k}(\omega,\vec{p}) - G^A_{\phi/j,k}(\omega,\vec{p}) \right) \, .
\end{align}
We compute the relevant 1PI vertex functions of Model~H via functional derivatives of the truncated effective average 

\noindent
MSR action (evaluated at vanishing field expectation values),
\begin{subequations}
\begin{align}
    \Gamma_k^{\tilde{\Phi}\phi j_l}(p,q,r) &= -g\,\frac{r^0 \, (\mathcal{T}_{\vec{r}} \vec{p})_l}{\eta_k \vec{r}^2 \, \sigma_k \vec{p}^2} \, ,\label{eq:GamPpj} \\
    \Gamma_k^{\tilde{J}_l \phi \phi}(p,q,r) &= g \, \frac{q^0 \, (\mathcal{T}_{\vec{p}} \vec{q})_l}{\eta_k \vec{p}^2\, \sigma_k \vec{q}^2} + g \,\frac{r^0 \, (\mathcal{T}_{\vec{p}}\vec{r})_l}{\eta_k \vec{p}^2\, \sigma_k \vec{r}^2} \, , \label{eq:GamJpp} \\
    \Gamma_k^{\tilde{\Phi}\tilde{\Phi}\phi\phi}(p,q,r,s) &= \frac{2ig^2 T}{\sigma_k \vec{p}^2\,\sigma_k \vec{q}^2} \, \left( \frac{\mathcal{T}_{lm}(\vec{p}+\vec{r})}{\eta_k(\vec{p}+\vec{r})^2} + \frac{\mathcal{T}_{lm}(\vec{q}+\vec{r})}{\eta_k(\vec{q}+\vec{r})^2} \right) r_l s_m \, ,\label{eq:GamPPpp} \\
    \Gamma_k^{\tilde{J}_l \tilde{J}_m \phi \phi}(p,q,r,s) &= 2i g^2 T \, \frac{(\mathcal{T}_{\vec{p}}(\vec{p}+\vec{r}))_l \, (\mathcal{T}_{\vec{q}}(\vec{q}+\vec{s}))_m}{\eta_k \vec{p}^2 \, \eta_k \vec{q}^2 \, \sigma_k (\vec{p}+\vec{r})^2} + 2i g^2 T \, \frac{(\mathcal{T}_{\vec{p}}(\vec{p}+\vec{s}))_{l}  \, (\mathcal{T}_{\vec{q}}(\vec{q}+\vec{r}))_{m}}{\eta_k \vec{p}^2 \, \eta_k \vec{q}^2 \, \sigma_k(\vec{q}+\vec{r})^2} \, , \label{eq:GamJJpp}
\end{align}\label{eq:ModHVertFncs}%
\end{subequations}
with the notation $(\mathcal{T}_{\vec{p}}\vec{q})_l \equiv \mathcal{T}_{lm}(\vec{p}) q_m$, i.e.~$\mathcal{T}_{\vec{p}} \vec{q}$ denotes the component of $\vec{q}$ that is transverse to $\vec{p}$.

Since the symmetry of thermal equilibrium is preserved during the FRG flow, the flow of the LGW free energy \eqref{eq:freeEnergyPhi} decouples from the rest \cite{Roth:2024rbi} and satisfies the standard $d$-dimensional Euclidean flow equation \cite{Wetterich:1992yh}. Hence, the flow of the static couplings $m^2_k$ and $\lambda_k$ assume standard forms and can be investigated independently. As mentioned in the main text, we employ the optimized regulator 
$R_k^{\phi}(\vec{p}) =  (k^2-\vec{p}^{\,2})\theta(k^2-\vec{p}^{\,2})$
for the order parameter, and set the regulator for the momentum density $R_k^j(\vec{p}) = 0$ to zero. With this regulator choice the flow of the dimensionless static couplings
$\bar{m}^2_k \equiv k^{-2} m_k^2$ and $\bar{\lambda}_k \equiv k^{d-4} \lambda_k$ is given by (here for arbitrary $N$ to cover also the case of $O(N)$ Model~G) 
\begin{align}
    k\partial_k \bar{m}^2_k &= -2\bar{m}_k^2 - \frac{  \Omega_d}{(2\pi)^d)} \, \frac{N+2}{3N}  \, \frac{\bar{\lambda}_k}{(1+\bar{m}_k^2)^2} \, , \label{eq:dm2dk} \\
    k\partial_k \bar{\lambda}_k &= (d-4)\bar{\lambda}_k + \frac{\Omega_d}{(2\pi)^d} \, \frac{2(N+8)}{3N} \, \frac{\bar{\lambda}_k^2}{(1+\bar{m}_k^2)^3} \, ,\label{eq:dlamdk}
\end{align}
where the case $N=1$ corresponds the $Z_2$ symmetry of Model~H.

Since the temporal gauge symmetry is preserved, corresponding Ward identities of the effective average MSR action imply that the mode coupling $g$ is protected from renormalization. This was shown explicitly for Model~G in Ref.~\cite{Roth:2024rbi}, and the proof directly translates to Model~H as well.
We project the real-time FRG flow onto the kinetic coefficients $\sigma_k$ and $\eta_k$ by differentiating Eqs.~\eqref{eq:defLambdak} and \eqref{eq:defEtak} in the main text with respect to the FRG scale~$k$, which results in
\begin{align}
    \partial_k \sigma_k &= -\frac{\sigma_k^2}{2iT} \lim_{\vec{p}\to 0} \vec{p}^2 \lim_{\omega\to 0} \frac{\delta^2 \partial_k \Gamma_k}{\delta \tilde{\Phi}(-p)\delta \tilde{\Phi}(p)} \bigg\rvert_{\text{0}} \, ,\label{eq:dSigmadkFormal} \\
    \partial_k \eta_k &= -\frac{\eta_k^2}{2iT} \lim_{\vec{p}\to 0} \vec{p}^2 \lim_{\omega\to 0} \frac{\mathcal{T}_{lm}(\vec{p})}{d-1} \frac{\delta^2 \partial_k \Gamma_k}{\delta \tilde{J}_l(-p)\delta \tilde{J}_m(p)} \bigg\rvert_{\text{0}} \, . \label{eq:dEtadkFormal}
\end{align}
The corresponding second functional derivatives in \eqref{eq:dSigmadkFormal} and \eqref{eq:dEtadkFormal} can be expressed as a sum over 1-loop Feynman diagrams. Inserting the 1PI vertex functions \eqref{eq:GamPpj}--\eqref{eq:GamJJpp}, evaluating the frequency integrals via the residue theorem and the momentum integrals analytically (which is possible due to the choice of the optimized regulator), we obtain
\begin{align}
    \partial_k \sigma_k&=\frac{2g^2 \Omega_d k^{d-1} T}{(2\pi)^d} \, \frac{d-1}{d-2} \, \frac{1}{\eta_k} \left( \frac{ \sigma_k^2}{(\eta_k/\rho+\sigma_k (k^2 +m_k^2))^2} -\frac{1}{(k^2 +m_k^2)^2} \right) \, , \label{eq:dsigmadk}  \\
    \partial_k \eta_k &=-\frac{g^2 \Omega_d k^{d+1} T}{(2\pi)^d \,(2+d)\sigma_k (k^2+m_k^2)^3 } \, . \label{eq:detadk}
\end{align}
The flow equations for the kinetic coefficients $\Gamma_k^{\phi}$ and $\gamma_k$ in $O(N)$ Model~G can be obtained from Ref.~\cite{Roth:2024rbi} by neglecting the non-trivial momentum dependence of $\gamma_{n,k}(\vec{p}) \approx \gamma_k \vec{p}^{\,2}$ considered therein. Again using the optimized regulator $R_k^{\phi}(\vec{p})$ for the order parameter as above, and setting the regulator $R_k^{n}(\vec{p}) = 0$ for the charge densities $n_{ab}$ to zero. With these choices, the momentum integrals on the right-hand sides of the flow equations for the kinetic coefficients $\Gamma_k^{\phi}$ and $\gamma_k$ can be solved analytically, and the result is
\begin{align}
    \partial_k \Gamma_k^{\phi} &= \frac{g^2 \, (N-1)  \,d \,\Omega _d k^{d-1} T}{(2\pi)^{d} \, \left(k^2+m_k^2\right) \, \gamma_k} \Bigg\{\frac{ \Gamma _k^{\phi } 
   }{k^2\gamma _k/\chi+  \Gamma _k^{\phi
   } ( k^2+ m_k^2 ) }  -\frac{2+(d-4) \; {}_2F_1\left(1,\frac{d-2}{2};\frac{d}{2};-\frac{k^2 \gamma _k/\chi}{\Gamma _k^{\phi
   } (k^2+m_k^2 )}\right)}{(d-2)\,\left(k^2+m_k^2\right)}\Bigg\}  \, , \label{Eq:dGammadk} \\
   \partial_k \gamma_k &= -\frac{2 g^2 \Omega _d k^{d+1} T}{(2\pi)^d \, \Gamma _k^{\phi } \,  \left(k^2+m_k^2\right)^3}\,, \label{Eq:dgammadk}
\end{align}
with the hypergeometric function ${}_2F_1$ in \eqref{Eq:dGammadk}, and with the volume factor $\Omega_d \equiv 2\pi^{d/2}/(\Gamma(d/2)\,d)$ in both cases. 

Specifically in case of Model~H we also consider the more sophisticated LPA' truncation of the free energy. This is possible here because an expansion around a non-vanishing field expectation value only affects the propagators \eqref{eq:ModHGRandGA}, but not the 1PI vertex functions \eqref{eq:ModHVertFncs} (in contrast to Model~G \cite{Roth:2024rbi}). In LPA' one takes into account the full effective potential $U_k(\rho)$ and a non-trivial wave function renormalization factor $Z_k^{\perp}$, but assumes that the latter is field-independent, approximated by its value at the some (possibly $k$-dependent) expansion point $\rho_{0,k}$. More specifically, the truncation of the free energy \eqref{eq:freeEnergyPhi} is given by (with the field invariant $\rho \equiv \phi_a \phi_a$ and for arbitrary $d$ and~$N$)
\begin{align}
    F_k =\!\int \! d^d x \left\{ \frac{Z_k^{\perp}}{2}(\partial^i \phi_a)(\partial^i \phi_a)+U_k(\rho) \right\} 
    \label{eq:freeEnergyPhiLPA'}
\end{align}
We also introduce a factor of $Z_k^{\perp}$ in the optimized regulator, which becomes $R_k^{\phi}(\vec{p}) = Z_k^{\perp} (k^2-\vec{p}^{\,2})\theta(k^2-\vec{p}^{\,2})$.
The flow equations for the effective potential $U_k(\rho)$ and the wave function renormalization factor $Z_k^{\perp}$ can be obtained using appropriate projections of the Wetterich equation \cite{Wetterich:1992yh}. For the dimensionless effective potential $\bar{U}_k(\bar{\rho})$, defined by $k^{d} T \,\bar{U}_k(\bar{\rho}) \equiv U_k(\rho)$ with $\rho \equiv k^{d-2} T \bar{\rho}/Z_k^{\perp}$, one obtains the well-known result~\cite{Berges:2000ew}
\begin{align}
    k\partial_k \bar{U}_k(\bar{\rho}) = -d\,\bar{U}_k(\bar{\rho}) + (d-2+\eta_k^{\perp}) \, \bar{U}_k'(\bar{\rho}) + \frac{\Omega_d}{(2\pi)^d} \left(1-\frac{\eta_k^{\perp}}{2+d}\right) \frac{1}{1+2\,\bar{U}'_k(\bar{\rho})+4\bar{\rho}\,\bar{U}_k''(\bar{\rho})} \,, \label{eq:dkUkLPA'}
\end{align}
where the $k$-derivative is taken at fixed $\bar{\rho}$.

The flow of the wave function renormalization factor $Z_k^{\perp}$ at the scale-dependent minimum $\bar{\rho}_{0,k}$ of the effective potential, $\bar{U}_k'(\bar{\rho}_{0,k}) = 0$, is encoded in the anomalous dimension $\eta^{\perp}_k \equiv -k\partial_k \log Z_k^{\perp}$ of the order-parameter field. Projecting the Wetterich equation onto the pion (Goldstone) channel for values $N \geq 2$ yields \cite{Tetradis:1993ts}
\begin{align}
    \eta^{\perp}_k = \frac{32\,\Omega_d\,\bar{\rho}_0\, \bar{U}_k''(\bar{\rho}_0)}{(2\pi)^d\,(1+4\bar{\rho}_0\,\bar{U}_k''(\bar{\rho}_0))^2} \,, \label{eq:anomDimLPA'}
\end{align}
By regarding this result as an analytic function of $N$ one can extend its range of validity by analytic continuation to $N=1$. It was found empirically that this procedure yields reasonable values for the anomalous dimension~\cite{Tetradis:1993ts} also in the $Z_2$ case, which is incentive enough for us to use the same procedure in the present work. In contrast, obtaining similarly accurate values for $\eta_{\perp}$ using a projection onto the longitudinal (sigma) channel requires the field dependence of $Z_k^{\perp}(\rho)$ \cite{Canet:2002gs}, which we leave for future work. 

In practice, we solve the LPA' system \eqref{eq:dkUkLPA'} and \eqref{eq:anomDimLPA'} for the potential $\bar{U}(\bar{\rho})$ and the anomalous dimension $\eta_{\perp}$ at the Wilson-Fisher fixed point $k\partial_k \bar{U}_k(\bar{\rho}) = 0$ numerically using a standard shooting method: First, requiring regularity at the origin demands the behaviour \[ \bar{U}(\bar{\rho}) \sim \left(1-\frac{\eta_{\perp}}{2+d}\right)\frac{\Omega_d}{(2\pi)^d\,d\,(1+2c_1)} + c_1\bar{\rho}+\cdots \hspace{0.3cm}\text{for $\bar{\rho} \to 0$} \,. \] Second, depending on the choices of $c_1$ and $\eta_{\perp}$, one generally encounters a singularity in the second derivative $\bar{U}''(\bar{\rho})$ already at a finite value of $\bar{\rho}$. The fixed-point solution is given for those values of $c_1$ and $\eta_{\perp}$ where the solution is regular for the entire domain $0\leq \bar{\rho} < \infty$. Hence, we vary the parameters $c_1$ and $\eta_{\perp}$ so as to maximize the domain where the solution is regular until we reach the desired accuracy for $c_1$ and $\eta_{\perp}$.

For a non-trivial $Z_k^{\perp} \neq 1$, the flow equations \eqref{eq:dsigmadk} and \eqref{eq:detadk} for the kinetic coefficients $\sigma_k$ and $\eta_k$ become
\begin{align}
    \partial_k \sigma_k&=\frac{2g^2 \Omega_d Z_k^{\perp} k^{d-1} T}{(2\pi)^d} \, \frac{(d-1)(d-\eta_{\perp})}{d(d-2)} \, \frac{1}{\eta_k} \left( \frac{ \sigma_k^2}{(\eta_k/\rho+\sigma_k (Z_k^{\perp} k^2 +m_k^2))^2} -\frac{1}{(Z_k^{\perp} k^2 +m_k^2)^2} \right) \, , \label{eq:dsigmadkEtap}  \\
    \partial_k \eta_k &=-\frac{g^2 \Omega_d (Z_k^{\perp})^2 k^{d+1} T}{(2\pi)^d \,(2+d)\sigma_k (Z_k^{\perp}k^2+m_k^2)^3 } \, . \label{eq:detadkEtap}
\end{align}
The definition \eqref{definitionwfH} of the dimensionless dynamic couplings $w_H$ and $f_H$ must be generalized as well according to
\begin{equation*}
    w_H \equiv \rho \, \frac{\sigma_k Z_k^{\perp}  k^2}{\eta_k} \,, \hspace{1.0cm} f_H \equiv \frac{d\,\Omega_d\, g^2 T}{(2\pi)^d} \, \frac{k^{d-4}}{Z_k^{\perp} \sigma_k \eta_k} \,.
\end{equation*}
By applying a $k$-derivative using \eqref{eq:dsigmadkEtap} and \eqref{eq:detadkEtap}, their flow can be computed to be
\begin{align}
    k\partial_k f_H &= (d-4+\eta_{\perp})f_H +  f_H^2 \bigg[\frac{1}{d(d+2)}\frac{1}{(1+\bar{m}^2)^3}-  \frac{2(d-1)(d-\eta_{\perp})}{d^2\,(d-2)}\bigg(\frac{1}{(1/w_H +(1+\bar{m}^2))^2} \!-\!\frac{1}{(1+\bar{m}^2)^2}\bigg) \bigg] \,, \label{eq:fHflowLPAp} \\
    k\partial_k w_H &= (2 -  \eta_{\perp})w_H + w_H f_H \bigg[ \frac{1}{d(d+2)}\frac{1}{(1+\bar{m}^2)^3} +\frac{2(d-1)(d-\eta_{\perp})}{d^2\,(d-2)}\bigg(\frac{1}{(1/w_H +(1+\bar{m}^2))^2} \!-\!\frac{1}{(1+\bar{m}^2)^2}\bigg) \bigg] \,,\label{eq:wHflowLPAp}
\end{align}
with the corresponding fixed point being located at
\begin{equation}
    w_H^*=0\,, \hspace{1.0cm}
    f_H^*=\frac{(d-2)(4-d-\eta_{\perp})(1+\bar{m}^2)^3}{\frac{2(d-1)(d-\eta_{\perp})}{d^2}(1+\bar{m}^2)+\frac{d-2}{d(d+2)}} \,. \label{ModelHf*Etap}
\end{equation}
The dynamic critical exponents $x_{\sigma}$ and $x_{\eta}$ are still given by evaluating $x_{\sigma} = -k\partial_k \log \sigma_k$ and $x_{\eta} = -k\partial_k \log \eta_k$ with \eqref{eq:dsigmadkEtap} and \eqref{eq:detadkEtap} at the fixed point \eqref{ModelHf*Etap}, which yields the generalization of \eqref{eq:xSigmaAndxEta} to a finite $\eta_{\perp} \neq 0$,
\begin{equation}
    x_{\sigma}= \frac{2(d-1)(d-\eta_{\perp})f_H^*}{d^2\,(d-2)(1+\bar{m}^2)^2} \,,\hspace{1.0cm}
    x_{\eta} = \frac{f_H^*}{d\,(d+2)(1+\bar{m}^2)^3} \,, \label{eq:xsigmaAndxetaEtap}
\end{equation}
with $f_H^*$ given by \eqref{ModelHf*Etap}.

\end{document}